# Observation of nonreciprocal magneto-optical scattering in nonencapsulated few-layered CrI$_3$

Kai Guo[†,‡], Zhen Liu[†,‡], Guangwei Hu[§], Zhongtai Shi[†], Yue Li[†], Linbo Zhang[†],

Haiyan Chen[†], Li Zhang[†], Peiheng Zhou[†], Haipeng Lu[†], Miao-Ling Lin[¶], Sizhao Liu[∥],

Yingchun Cheng[∥], Xue Lu Liu[4], Jianliang Xie[†], Lei Bi[†], Ping-Heng Tan[¶], Longjiang

Deng[†,*], Cheng-Wei Qiu[§,*] and Bo Peng[†,*]

[†] National Engineering Research Center of Electromagnetic Radiation Control Materials, School of Electronic Science and Engineering, University of Electronic Science and Technology of China, Chengdu 611731, China

[§] Department of Electrical and Computer Engineering, National University of Singapore, 117583, Singapore

[∥] Key Laboratory of Flexible Electronics & Institute of Advanced Materials, Jiangsu National Synergetic Innovation Center for Advanced Materials, Nanjing Tech University, Nanjing 211816, China

[¶] State Key Laboratory of Superlattices and Microstructures, Institute of Semiconductors, Chinese Academy of Sciences, P. O. Box 912, Beijing 100083, China

[‡] Kai Guo and Zhen Liu equally contribute to this work.

* To whom correspondence should be addressed. Email address:

bo_peng@uestc.edu.cn (B.P); chengwei.qiu@nus.edu.sg (C.W.Q); denglj@uestc.edu.cn (L.J.D)

**Abstract:** "Magneto-optical" effect refers to a rotation of polarization plane, which has been widely studied in traditional ferromagnetic metal and insulator films and scarcely in two-dimensional layered materials. Here we uncover a new nonreciprocal magneto-inelastic light scattering effect in ferromagnetic few-layer $CrI_3$. We observed a rotation of the polarization plane of inelastic light scattering between $-20^o$ and $+60^o$ that are tunable by an out-of-plane magnetic field from -2.5 to 2.5 T. It is experimentally observed that the degree of polarization can be magnetically manipulated between -20% and 85%. This work raises a new magneto-optical phenomenon and could create opportunities of applying 2D ferromagnetic materials in Raman lasing, topological photonics, and magneto-optical modulator for information transport and storage.

**INTRODUCTION**

Ferromagnetic ordering at the monolayer limit was believed to be prohibited at nonzero temperature according to the Mermin–Wagner theorem (*1*), however, recent discoveries of Ising ferromagnetism in monolayer $CrI_3$ and $Cr_2Ge_2Te_6$ prove that there remain exciting possibilities of magnetism at the atomic scale thickness in two-dimensional (2D) material (*2-3*). The rich physics of the magneto-optical effect and spin manipulation in 2D ferromagnetic van der Waals (vdW) material remains to be extensively explored (*4-7*). The long-range magnetic orders together with their rich electronic and optic properties in 2D layered materials could open up numerous opportunities for topological photonics, spintronics and information storage applications (*8-15*).

Atomically thin $CrI_3$ is a typical 2D Ising ferromagnetic material with a Curie temperature ($T_C$) of ~45 K (*3, 16*). The ferromagnetic nature originates from the $Cr^{3+}$ ions with electronic configuration $3d^3$; the magnetic moment is aligned in the out-of-plane direction perpendicular to the $CrI_3$ layer. Beyond the monolayer, the vdW heterostructures comprising layered 2D ferromagnetic materials have been studied as magnetic tunneling junction devices, showing the promise for next-generation information transfer and data storage technologies (*11-12, 17*). Moreover, the feasible approaches to switch the magnetic orders between ferromagnetic (FM) and antiferromagnetic (AFM) states through electric gating or electrostatic doping (*18-20*) and pressure (*21-22*) have been demonstrated, again providing more opportunities for the reliable 2D magnetic devices.

However, to date, the magneto-optical effect in layered 2D ferromagnetic materials is rarely explored except for Kerr effect (*23*). For example, magnetic fields tune inter-Landau-level excitations in resonance with the phonon by electron-phonon coupling, resulting in a pronounced anti-crossing behavior of the coupled phonon and electronic modes in nonmagnetic graphene (*24-25*). As a simple physical picture, the Lorentz oscillator mode gives the polarizability and permittivity of materials *via* the kinetic equation of motion; in the presence of an applied out-of-plane magnetic field ($\vec{B} \parallel \hat{z}$) that leads to a Lorentz force $eB\dot{r}$, the kinetic equation is modified as is given $m\ddot{r} + m\gamma\dot{r} + m\omega_0^2 r(\omega) + eB\dot{r} = -eE$. This indicates that the polarizability and permittivity can be controlled by magnetic field (*26-27*), thus suggesting dramatic inelastic phonon scattering behavior controllable with magnetic field. This phenomenon has rarely been reported until now. This missing scenario may hinder the full manipulation of photons, electrons and spins, especially, phonons in atomically thin magnetic materials as well as the exciting applications based on the magneto-optical effect.

Here we report a phenomenal effect of nonreciprocal magneto-optical phonon scattering. Specifically, the low-temperature polarization-angle-resolved confocal Raman system with the external magnetic field is employed to study the anisotropic inelastic phonon scattering. We observe that the polarization axis of the light scattered from 3L, 5L and bulk $CrI_3$ rotates from -15°, -8°, -20° to +40°, +35°, +60°, respectively, when an applied out-of-plane magnetic field changes from -2.5 to +2.5 T. The helicity parameter of circularly polarized phonon scattering light can be magnetically tuned between -20% and 85%, which are also nonlinear and nonreciprocal *vs* magnetic field.

Our measurements reveal that the ferromagnetic 2D layered CrI$_3$ is a unique magneto-optical material system that supports the seemingly controversial coexistence between the extremely large rotation angle and short optical path; and thereafter holds great promise for nonreciprocal magneto-optical devices, especially in Raman lasing toward photonic integrated circuit and on-chip devices (*28-31*).

**RESULTS AND DISCUSSION**

The CrI$_3$ bulk crystals were synthesized by chemical vapor transfer (CVT) method (*16*). Although, the bulk is monoclinic (space group C2/m, see Supplementary information Section 1) at room temperature, the few-layer and bulk CrI$_3$ is rhombohedral stacking order at 10 K (space group $R\bar{3}$, Fig. 1A), consistent with our previous reports (*32*). The trilayer (3L) and five-layer (5L) CrI$_3$ were firstly exfoliated from the bulk and then were directly transferred onto 285 nm SiO$_2$/Si substrates; the non-encapsulated CrI$_3$ were in-situ loaded into a closed cycle optical cryostat in glovebox. Fig. 1B shows the optical micrograph of a 3L CrI$_3$ flake, and the layer numbers of CrI$_3$ can be identified through the optical contrast (*3, 32*). The magnetic control of magneto-optical scattering effect is then studied through our low-temperature (10 K) angle-resolved Raman system with an out-of-plane external magnetic field (7 T), which have been excellently calibrated (see Methods and Supplementary information Section 2).

Fig. 1C displays the Raman features of 3L, 5L and bulk CrI$_3$ at 10 K in the absence of magnetic field. The features at ~128 and ~237 cm$^{-1}$ in 3L CrI$_3$ are assigned to the out-of-plane $A_g^3$ mode (*32-33*) (the alternate squeezing and spreading of iodine atoms in the out-of-plane direction), and the in-plane $E_g^4$ modes (the alternate displacement of Cr atoms) (*34-35*). We observed that the $E_g^4$ features are softened and shift to low frequency as the layer number increases, while the $A_g^3$ features remains unaffected. The frequency difference between $A_g^3$ and $E_g^4$ modes changes from 109.2, 108.6 to 107.6 cm$^{-1}$ in 3L, 5L and bulk CrI$_3$, respectively, indicating the feasibility of identifying the layer number through the frequency difference.

Importantly, the unambiguous nonreciprocal magneto-optical scattering effect is observed. Figures 1D-F show the polarization dependence of the $A_g^3$ and $E_g^4$ modes in 3L CrI$_3$ at 10 K excited by a linearly polarized 514 nm light with an incident polarization ($e_i$) along the x axis, defined as 0°. The polarization axis of scattering light ($e_s$) is rotated in the x-y plane with $\theta$ angle configuration. The Raman intensities of the $E_g^4$ mode almost remain constant in different scattering polarization, indicating that the in-plane $E_g^4$ vibration is isotropic. Furthermore, the isotropic behaviors are independent on the out-of-plane magnetic field (see Supplementary information Section 3-5). In contrast, the $A_g^3$ mode exhibits a pronounced anisotropic polarization dependence behavior with a two-fold rotation symmetric shape. At zero external magnetic fields, the scattering polarization axis of the $A_g^3$ mode is at 5° (see Fig. 1D and Supplementary information Section 3). Strikingly, when an external magnetic field is applied, the scattering polarization axis of the anisotropic $A_g^3$ mode is rotated to +40° (-15°) value for +2 T (-2 T). It is very interesting to note that the rotation changes its direction as well as the amplitude, too, when the magnetic field with the same magnitude switches the direction (see Supplementary information

Section 3).

To further verify such exotic magneto-optical effect, we measured the polarization-resolved Raman spectra of 5L and bulk $CrI_3$ under different magnetic fields. Figure 2A shows the polarization dependence of Raman intensities of $A_g^3$ mode of 5L $CrI_3$ in different magnetic field. Strikingly, the polarization axis is rotated from -8° to +35° as the magnetic field variate from -2.5 T to +2.5 T (see Fig. 2B and Supplementary information Section 4). The rotation direction of polarization axis is opposite when the direction of magnetic field is reversed, however, the absolute values of polarization rotation angles are not equal in the positive and negative magnetic field. Such nonreciprocal magneto-optical scattering behaviors are more explicit in bulk $CrI_3$. The polarization axis is slowly rotated to -20° as magnetic field increase from 0 T to -2.5 T, however, drastically rotated to +60° with increasing magnetic field from 0 T to +2.5 T (see Fig. 2B and Supplementary information Section 5). In stark contrast, using Si substrates as control samples, the scattering polarizations show no dependence on the magnetic field, although a two-fold anisotropic pattern is observed too (see Supplementary information Section 6). Therefore, it evidences that nonreciprocal magneto-optical scattering effect of anisotropic phonon mode takes place in 2D ferromagnetic $CrI_3$ in the presence of magnetic field.

The Raman scattering features of $A_g^3$ and $E_g^4$ mode of 3L, 5L and bulk $CrI_3$ at different magnetic field do not offer any detectable difference on the phonon frequency, line width and shape (see Supplementary information Section 7), indicating that the lattice parameters of CrI$_3$ are completely unrelated to the external magnetic field. Thus, the magneto-strictive strain effect, the coupling of phonon and Landau level, and phonon-plasmon coupling are not responsible for the nonreciprocal magneto-optical scattering effect (*24-25, 36*). The un-polarized Raman scattering intensities of $A_g^3$ and $E_g^4$ mode detected without the polarization analyzer are also independent on magnetic field (see Supplementary information Section 8), indicating that magnetic field does not influence the total scattering intensity. Thus, we can safely attribute the magneto-optical scattering effect to magnetic control of dipole moment of CrI$_3$ lattice vibration (Fig. 3A and B).

In ferromagnetic materials, the spin-orbital interaction and Heisenberg's exchange interaction lead to a strong effective magnetic field and lift the degeneracy of ground states and excited states together with the external magnetic field, thus resulting in nonreciprocal magneto-optical Faraday and Kerr effect of ferromagnets (*9-10, 37-38*). Generally, classic theories have been proposed to explain the magneto-optical Faraday and Kerr effect in ferromagnetic materials: (1) the electron dynamics theory from Lorentz model and Maxwell's equations; (2) the macroscopic theory on the basis of the dielectric tensor theory in the constitutive relations solving the Maxwell's equations of electromagnetic wave. The electron dynamics theory from Lorentz model demonstrates that the polarizability tensor ($\alpha_K$) is strongly dependent on magnetic field (see Supplementary information Section 9). From the macroscopic descriptions of scattering, Raman inelastic scattering intensity is intimately connected to the differential scattering cross sections and polarizability tensor, which is given by

$$I = \left[ \frac{\langle Q_{K0} \rangle (\omega_0 - \omega_k)^4}{32\pi c^3} \sin^2 \varphi \right] |\alpha_K \cdot E|^2 \quad (1)$$

Onsager's relation has proposed that the symmetry of off-diagonal components of the polarizability tensor is broken by a magnetic field due to time inversion symmetry breaking (*39*), thus $\alpha_{xy} \neq \alpha_{yx}$ and the reciprocity is broken (see Supplementary information Section 9 and 10). The nonreciprocity increases as the ratio of the off-diagonal component to the diagonal component increases. The scattering intensity is maximum only when the dipole moment vector is perpendicular to scattering direction ($\varphi=90°$), thus the rotation angle of polarization axis ($\theta$, Fig. 2C) is strongly dependent on $a_{yx}/a_{xx}$. That is to say, the magneto-optical scattering effect originates from the rotation of the dipole moment vector through tuning the polarizability tensor (Fig. 3A and B).

The few- and mono-layer CrI$_3$ have shown Kerr effect and magnetic circular dichroism in the NIR-visible light region (*3, 14, 20*); the ratio of the off-diagonal to diagonal components of the permittivity tensor of few-layer CrI$_3$ is up to ~5%, representing at least one order of magnitude larger than classic ferromagnetic insulator CeYIG and BiYIG (see Supplementary information Section 11) (*40*), thus leading to the giant magneto-optical scattering angle up to approximately 2 orders of magnitude larger than that from the magneto-optical Kerr effect (*23*). Moreover, the CrI$_3$ flakes may be ideal 2D magneto-optical materials and open the door toward nonreciprocal photonic integrated device at optical frequencies (*28-29, 31*). To avoid complex mathematic derivation and focus on the physical essence of magneto-optical scattering effect, the polarizability tensor of $A_g^3$ mode in the presence of magnetic field is simplified to a new Hermitian matrix with magnetic-dependent complex diagonal and off-diagonal components (see Supplementary information Section 9 and 10):

$$\alpha_K^L(A_g(B)) = \frac{e}{\omega}\begin{bmatrix} -\frac{iaB_0}{B_0^2+B^2} & \frac{bB}{B_0^2+B^2} & 0 \\ -\frac{bB}{B_0^2+B^2} & -\frac{iaB_0}{B_0^2+B^2} & 0 \\ 0 & 0 & \alpha_{zz} \end{bmatrix}$$

The corresponding magneto-optical scattering intensity of $A_g^3$ mode as a function of scattering angle $\theta$ is given by

$$I(A_g(B)) \propto \left| -\frac{e}{\omega}\frac{(Bb\sin\theta - ha\cos\theta) + iga\cos\theta}{(g+ih)^2 + B^2} \right|^2 \quad (2)$$

where $a$ and $b$ are the lattice symmetry-restricted elements, and $B_0 = \frac{m(-\omega^2 + i\gamma\omega + \omega_0^2)}{ie\omega} = \frac{m\gamma}{e} + i\frac{m(\omega^2 - \omega_0^2)}{e\omega} = g + ih$ determines the rotation of polarization aixs and ellipticity of the polarized scattering light. All experimental results of $A_g^3$ modes from 3L, 5L and bulk CrI$_3$ are together fitted by Eq. 2, in which the experimental and calculated results are in good agreement (Fig. 3C and Supplementary information Fig. S9A-C). More importantly, the electron moving in an out-of-plane magnetic field experiences an in-plane Lorentz force, which increase the natural frequency ($\omega_0$) in an negative magnetic field. Consequently, the damping constant $\gamma$ and $B_0$ increase in negative magnetic field (Fig. S9D), while Lorentz force slows down $\omega_0$ and deminishes $B_0$ in a positive magnetic field (Fig. S9E). Therefore, the diagonal and off-diagonal components of polarizability tensor can be controlled by magnetic field and display different values in negative and positive magnetic field, enabling the non-reciprocal magneto-optical scattering effect (see Supplementary information Section 10, Fig. S10). Figure 3D shows that the rotation angle of polarization axis increases as the $g$ decreases. Strikingly, the non-reciprocal magneto-optical scattering behaviors are predicted by the proposed Lorentz-modified Eq.(2) from the electron dynamics theory. The polarization angle is ~60° in a magnetic field of 2.5 T with $g$=3.8. On the contrary, the angle changes to approximately -20° when the magnetic field switches to -2.5 T with $g$=4.3. The experimental results in Fig. 2B are in agreement with the calculation. Thus, the non-reciprocal magneto-optical scattering effect arises from magnetic-field-dependence of $B_0$, which tune both complex diagonal components and off-diagonal components of polarizability tensor (see Supplementary information Section 9 and 10) (27). An intrinsic magnetic field of $CrI_3$ flakes below $T_c$ enhances the anisotropic scattering effect, resulting in a non-zero scattering angle at 0 T (3), as shown in Fig. 2B and Supplementary information Section 3-5. The magnetic-field-dependent behaviors of inelastic scattering polarization angle mainly arise from the Lorentz force. The exotic magneto-biased magneto-inelastic scattering effect, in which the polarization angle value is asymmetric and nonreciprocal in the positive and negative magnetic field (Fig. 2B), deserves future theoretical investigation. It may lead to the better understanding of the fundamental physics of the anomalous nonreciprocal magneto-optical scattering effect in 2D ferromagnetic layered materials. The nonreciprocal magneto-optical scattering effect is unambiguously observed in ferromagnetic few-layer $CrI_3$, which uncovers a nonreciprocal magneto-optical effect, besides Faraday effect, Kerr effect and magnetic circular dichroism. Our observation may stimulate both experimental and theoretical investigations of magneto-optical scattering effect.

The Raman intensity of $E_g^4$ mode at zero magnetic field is given by $I(E_g) \propto c^2 + d^2$ (see Supplementary information Section 12), which confirms its isotropic behaviors. The corresponding magneto-optical Raman scattering intensity of $E_g^4$ mode in an out-of-plane magnetic field is $I(E_g(B)) \propto |\frac{B_0}{B_0^2 + B^2} c|^2 + |\frac{B}{B_0^2 + B^2} d|^2$ (see Supplementary information Section 13). The in-plane $E_g^4$ mode intensities are still independent on the polarization angle, which remains isotropic under magnetic field. Therefore, the $E_g^4$ mode intensity remains almost constant and isotropic in different magnetic field (see Supplementary information Section 3-5).

Furthermore, we demonstrate a magnetic control of degree of polarization, which allows us to identify and controllably access the parallel and perpendicular polarization configuration. Fig. 4A and B show the magneto-optical Raman scattering intensity in the two-polarization configuration as a function of magnetic field. The Raman scattering intensity in cross-polarization configuration increase as the magnetic field increase, whereas the intensity in parallel-polarization configuration decreases. Remarkably, the degree of polarization (DOP) defined as $\eta = (I_{//} - I_{\perp})/(I_{//} + I_{\perp})$ is up to ~85%, and can be tuned monotonically from -20% to 85% when the magnetic field decreases from ±2.5 T to 0 (Fig. 4C). These results raise the feasibility of applying 2D ferromagnetic few-layer $CrI_3$ for information encoding and data storage.

**CONCLUSION**

In summary, a giant nonreciprocal magneto-optical scattering effect is observed in ferromagnetic few-layer $CrI_3$. The magnetic control of anisotropic Raman scattering, in which the polarization angle, ellipticity and DOP are magnetic-field dependent, has been demonstrated. The non-reciprocal magneto-optical scattering originates from magnetically control of the natural frequency by Lorentz force, leading to non-reciprocally magnetic manipulation of diagonal components and off-diagonal components of polarizability tensor. Our results demonstrate unique potential of 2D ferromagnets for magnetically controlled information encoding and Raman lasing applications.

**MATERIALS AND METHODS**

**$CrI_3$ synthesis and preparation:** The $CrI_3$ bulk crystals were synthesized by chemical vapor transport method. Chromium powder (99.99%, Alfa Aesar) and anhydrous iodine particles (99.99%, Alfa Aesar) were mixed in stoichiometric. About 1 g of the mixture was loaded in the ampoule (16 mm of inner diameter, 20 mm of outer diameter and 200 mm in length). The ampoule with mixture was placed into liquid nitrogen to prevent the sublimation of iodine particles and then evacuated to a pressure of approximately $10^{-3}$ Pa. The ampoule was sealed and placed into a two-zone furnace with temperature gradient of 650 ℃ to 530 ℃ for 7 days. The $CrI_3$ crystal with shiny and black was obtained at the sink region of ampoule. The crystal was stored in a nitrogen atmosphere and anhydrous conditions.

The few-trilayer $CrI_3$ were mechanically exfoliated from bulk crystal onto PDMS

films and then directly transferred onto SiO$_2$/Si substrates, which were in-suit loaded into the cold head for optical measurements in glovebox.

**Raman measurements:** The Raman signals were recorded by a Witec Alpha 300R Plus confocal Raman microscope, which is coupled with a closed cycle He optical cryostat (10 K) and a superconducting magnet. A long work distance 50× objective (NA = 0.45) was used for the Raman measurement at 10 K and magnetic field. The Raman signals were first collected by a photonic crystal fiber and then coupled into the spectrometer with 1800 g/mm grating. The polarization-resolved Raman spectra were obtained by rotating the polarization of analyzer which was put before the photonic crystal fiber. The power of the excitation laser at 514 nm was measured to about 2 mW and the typical integration time is 20 s. The magneto-optical system is checked by angle-resolved polarized Raman spectra using highly ordered pyrolytic graphite (HOPG) as reference samples; the G mode intensities of HOPG are isotropic on its basal plane, verifying that our optical system is excellently calibrated (see Supplementary information Section 2).

**SUPPLEMENTARY MATERIALS**

Supplementary material for this article is available at http://advances.sciencemag.org

32. K. Guo, B. Deng, Z. Liu, C. Gao, Z. Shi, L. Bi, L. Zhang, H. Lu, P. Zhou, L. Zhang, Y. Cheng, B. Peng, Layer dependence of stacking order in nonencapsulated few-layer $CrI_3$. *Sci. China. Mater.* **63**, 413-420 (2020)

33. S. Djurdjić-Mijin, A. Šolajić, J. Pešić, M. Šćepanović, Y. Liu, A. Baum, C. Petrovic, N. Lazarević, Z. V. Popović, Lattice dynamics and phase transition in $CrI_3$ single crystals. *Phys. Rev. B* **98**, 104307 (2018)

34. D. Shcherbakov, P. Stepanov, D. Weber, Y. Wang, J. Hu, Y. Zhu, K. Watanabe, T. Taniguchi, Z. Mao, W. Windl, J. Goldberger, M. Bockrath, C. N. Lau, Raman spectroscopy, photocatalytic degradation, and stabilization of atomically thin chromium tri-iodide. *Nano Lett.* **18,** 4214-4219 (2018)

35. L. Webster, L. Liang, J.-A. Yan, Distinct spin–lattice and spin–phonon interactions in monolayer magnetic $CrI_3$. *Phys. Chem. Chem. Phys.* **20**, 23546-23555 (2018)

36. W. Zhao, Q. Wu, Q. Hao, J. Wang, M. Li, Y. Zhang, K. Bi, Y. Chen, Z. Ni, Plasmon-phonon coupling in monolayer $WS_2$. *Appl. Phys. Lett.* **108**, 131903 (2016)

37. J. F. Dillon, H. Kamimura, J. P. Remeika, Magneto-optical properties of ferromagnetic chromium trihalides. *J. Phys. Chem. Solids* **27**, 1531-1549 (1966)

38. J. Suits, Faraday and kerr effects in magnetic compounds. *IEEE Trans. Magn.* **8**, 95-105 (1972)

39. Z. Q. Qiu, S. D. Bader, Surface magneto-optic kerr effect. *Rev. Sci. Instrum.* **71**, 1243-1255 (2000)

40. M. C. Onbasli, L. Beran, M. Zahradník, M. Kučera, R. Antoš, J. Mistrík, G. F. Dionne, M. Veis, C. A. Ross, Optical and magneto-optical behavior of cerium yttrium iron garnet thin films at wavelengths of 200-1770 nm. *Sci. Rep.* **6**, 23640 (2016)
**Acknowledgements**

**Funding:** We acknowledge the financial support from National Science Foundation of China (51602040, 51872039), Science and Technology Program of Sichuan (M112018JY0025), Scientific Research Foundation for New Teachers of UESTC
18

(A03013023601007) and the Ministry of Science and Technology of China MOST (No. 2016YFA0300802). C.-W.Q. acknowledges financial support from A*STAR Pharos Program (grant number 15270 00014, with project number R-263-000-B91-305).

**Author contributions:** B.P., L.J.D. and C.W.Q. developed the concept, designed the experiment, and prepared the manuscript. Y.C.C., S.Z.L. synthesized the $CrI_3$ crystal. K.G., Z.L., Z.T.S., and Y.L. prepared the $CrI_3$ samples, performed the polarization resolved Raman measurements and study temperature dependence. P.H.T., L.B., G.W.H., C.W.Q., M.L.L., X.L.L, L.B.Z., H.Y.C., L.Z., H.P.L., P.H.Z., J.L.X. contributed to the discussion of mechanism of magneto-optical scattering effect. All authors contributed to the editing and revision of the manuscript.

**Competing interests:** The authors declare no competing interests.

**Data and materials availability:** All data needed to evaluate the conclusions in the paper are present in the paper and/or the Supplementary Materials. Additional data related to this paper may be requested from the corresponding authors.

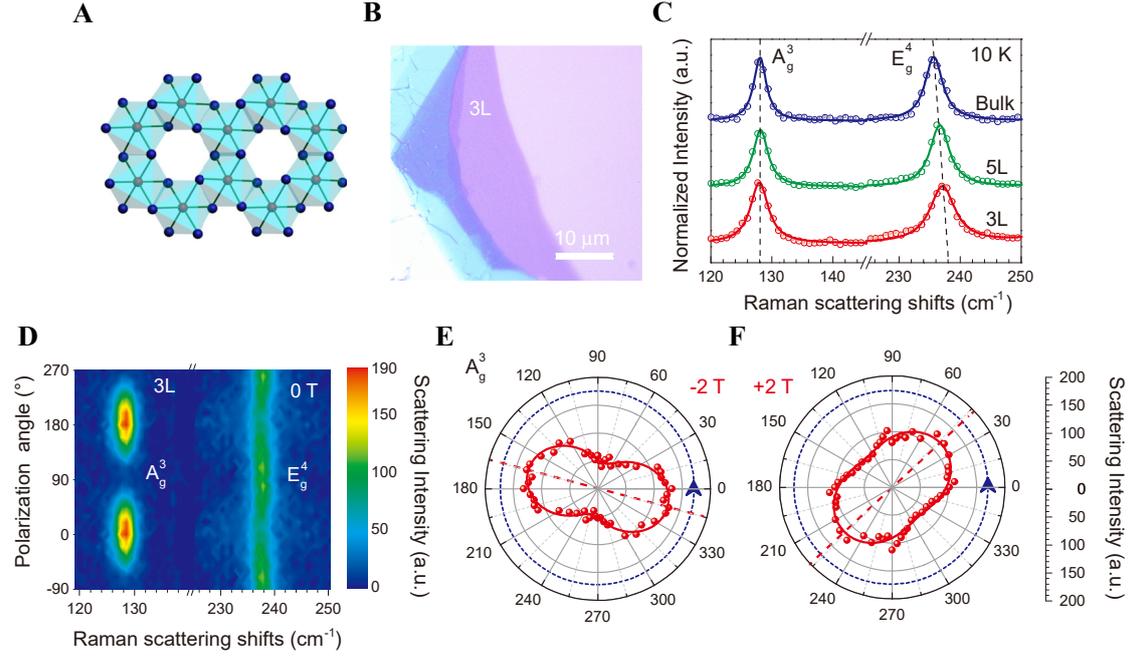

**Fig. 1. Structure of CrI$_3$.** (**A**) Schematic of monolayer CrI$_3$. Red and blue balls represent Cr and I atoms, respectively. The low temperature crystal structure of CrI$_3$ is rhombohedral (space group $R\bar{3}$), in which Cr$^{3+}$ ions are coordinated by six non-magnetic I$^-$ ions to form an octahedral geometry, which further share edges to build a honeycomb network. (**B**) The optical images of few-layered CrI$_3$ with an optical contrast of approximately -0.1, which determine the layer number to be 3. (**C**) Raman spectra of 3L, 5L and bulk CrI$_3$ at 10 K. The $E_g^4$ mode shift to low Raman frequency as the layer number increases. (**D**) Two-dimensional maps of phonon scattering spectra of 3L CrI$_3$ at 0 T. (**E, F**) Raman scattering intensity of $A_g^3$ and $E_g^4$ mode of 3L CrI$_3$ as a function of polarization angle under a magnetic field of -2 and +2 T at 10 K. Red dashed line represents the polarization axis of anisotropic scattering, which is rotated from -15° to 40° by the magnetic field.

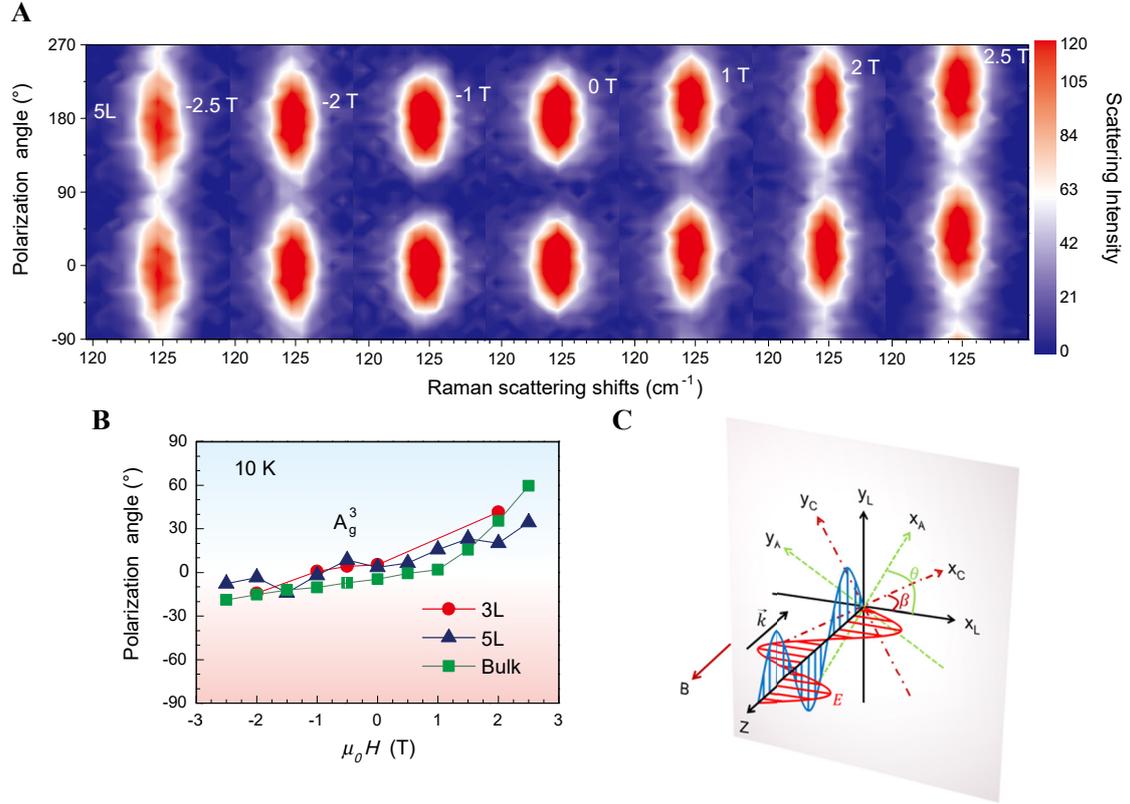

**Fig. 2. Magnetic control of anisotropic scattering.** **(A)** Two-dimensional maps of magneto-optical Raman scattering spectra of 5L CrI$_3$ as a function of magnetic field. The polarization plane is rotating as the magnetic field increase. **(B)** Polarization angle of 3L, 5L and bulk CrI$_3$ with an asymmetric variation as a function of magnetic field, indicating a nonreciprocal magnetic-biased control behavior. **(C)** Schematic picture of polarization configuration. $\beta$ is the the angle between laboratory coordinates ($x_L$, $y_L$, $z_L$) and crystal coordinates ($x_C$, $y_C$, $z_C$). $\theta$ is the scattering angle in the laboratory coordinates.

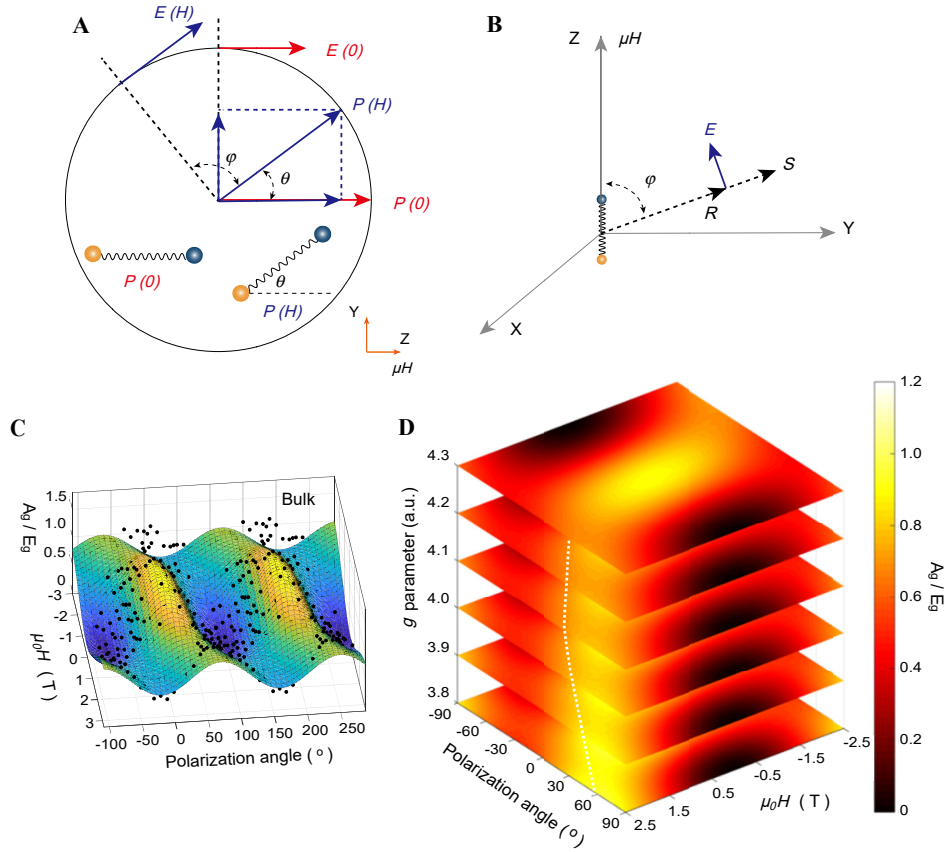

**Fig. 3. Origin of magneto-optical scattering effect. (A, B)** The vector of induced dipole moment ($P(0)$ and $P(H)$) is rotated by magnetic field from 0° to $\theta$, leading to that the direction of the maximum differential scattering cross sections ($E(0)$ and $E(H)$) rotates an angle of $\theta$. The $\varphi$ is the angle between the dipole moment ($P$) and the scattering direction ($R$); the $S$ represents the Poynting vector; the $E$ is the electric filed vector which is perpendicular to $R$ in the plane comprising $R$ and $P$. The $E(0)$ and $E(H)$ is maximized when $\varphi=90°$. **(C)** All experimental and corresponding calculation (three-dimensional color distributions, $g=4$, Fig. S9C) results of the $A_g^3$ mode of bulk CrI$_3$ vs magnetic field and polarization angle, which are well consistent. The intensity ratio of $A_g^3$ to $E_g^4$ mode is used to eliminate the system disturbance in the fitting. **(D)** Corresponding calculation results of $A_g^3$ mode of bulk CrI$_3$ as a function of parameter $g$, magnetic field and polarization angle.

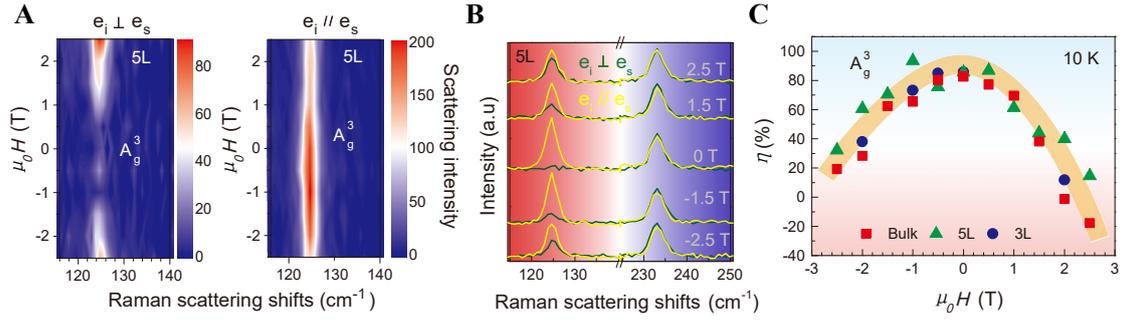

**Fig. 4. Experimental results on magnetic control of degree of polarization. (A, B)** Raman scattering spectra of 5L CrI$_3$ in the parallel and perpendicular polarization configuration as a function of magnetic field. The $A_g^3$ mode intensity can be controlled by a magnetic field and display an opposite magnetic-field dependence in perpendicular and parallel configuration. **(C)** Degree of polarization as a function of magnetic field.